\documentclass[12pt]{article}
\usepackage{amssymb}
\usepackage{amsmath}
\usepackage{hyperref}
\begin{document}
\title{\bf   Holographic Aspects of Non-minimal $R^3 F^2 $ Black Brane in an EFT Framework}
\author{ 
      Mehdi Sadeghi\thanks{Email:  mehdi.sadeghi@abru.ac.ir}\hspace{2mm}\\
{\small {\em Department of Physics, Faculty of Basic Sciences,}}\\
        {\small {\em Ayatollah Boroujerdi University, Boroujerd, Iran}}
       }
\date{\today}
\maketitle

\abstract{This work investigates a modified theory of gravity where the Einstein-Hilbert action, including a cosmological constant, is non-minimally coupled to a Yang-Mills field via an \(R^3 F_{\mu \alpha}^{(a)} F^{(a)\mu \alpha}\) interaction term. We treat this coupling as the leading higher-derivative correction in a low-energy effective field theory (EFT) deformation of the standard Einstein-Yang-Mills theory. We derive a black brane solution for this model, accurate to the first order in the EFT coupling parameter \(q_2\), and specify the regime of validity \(\frac{|q_2|}{L^6} \ll 1\). Using gauge/gravity duality techniques, we then compute two key holographic transport coefficients: the color non-abelian direct current (DC) conductivity and the ratio of shear viscosity to entropy density. Our analysis reveals that both transport coefficients are modified by the non-minimal coupling, with the conductivity bound violated for positive \(q_2\) and the Kovtun-Son-Starinets (KSS) bound for shear viscosity violated for negative \(q_2\). The results are interpreted within the EFT framework, and possible constraints on the sign of \(q_2\) from stability and causality are discussed. In the limit where the non-minimal coupling vanishes, our results consistently reduce to those of the standard Yang-Mills Schwarzschild Anti-de Sitter (AdS) black brane.}\\

\noindent PACS numbers: 11.10.Jj, 11.10.Wx, 11.15.Pg, 11.25.Tq\\

\noindent \textbf{Keywords:}  AdS/CFT duality, DC Conductivity, Black brane, Shear viscosity, Entropy density, Effective field theory
\section{Introduction} \label{intro}

Black holes, as solutions to Einstein's field equations characterized by their event horizons, represent one of the most intriguing predictions of general relativity. The classical notion that nothing escapes the event horizon was revolutionized by the discovery of Hawking radiation, providing a quantum mechanical window into their thermodynamic properties \cite{Hawking1983}. Understanding the quantum and thermodynamic aspects of black holes remains a central pursuit in theoretical physics.

A profound development in this field is the Anti-de Sitter/Conformal Field Theory (AdS/CFT) correspondence \cite{Maldacena}-\cite{Aharony}. This duality posits an equivalence between a gravitational theory in an asymptotically AdS spacetime and a conformal field theory (CFT) residing on its boundary. This framework provides a powerful tool for analyzing strongly coupled quantum field theories by mapping them to a classical gravity problem. In particular, the long-wavelength, low-frequency limit of this duality leads to the fluid-gravity correspondence \cite{Son}-\cite{Bhattacharyya}, where the boundary theory is described by hydrodynamics. Within this framework, transport coefficients of the boundary fluid, such as conductivity and shear viscosity, can be computed from the gravitational dual using methods like the Green-Kubo formula \cite{Son}.

A compelling direction in gravitational model-building involves non-minimal couplings between curvature invariants and matter fields. Such couplings, which modify the standard Einstein-Hilbert action, can offer alternative explanations for cosmological phenomena like inflation and late-time acceleration \cite{Bamba:2008ja}. Among the various possibilities, coupling the Ricci scalar to the Yang-Mills field strength invariant, $F^{(a)}_{\mu \alpha }F^{(a)\mu \alpha}$, represents a particularly rich class of non-minimal models \cite{Balakin:2015gpq, Edery:2018jyp}. These theories have been shown to admit exact solutions for stars \cite{Horndeski:1978ca, Mueller-Hoissen:1988cpx}, wormholes \cite{Balakin:2007xq, Balakin:2010ar}, and black holes \cite{Balakin:2007am, Balakin:2015oea}, often utilizing the Wu-Yang ansatz.

While previous studies have explored lower-order couplings like $RF^2$ \cite{Sadeghi:2023hxd} and $R_{\mu\alpha\nu\beta}F^{(a)\mu\alpha}F^{(a)\nu\beta}$ \cite{Sadeghi:2022bsh}, the present work extends this program to the next order by investigating the holographic implications of an $R^3F^2$ coupling. From an effective field theory (EFT) perspective, such higher-order curvature terms are naturally expected as the leading corrections to the low-energy action of a more fundamental theory. We therefore treat the $R^3F^2$ interaction as an EFT deformation, characterized by a dimensionful coupling $q_2$ (with $[q_2]=[L]^6$) that is suppressed by an ultraviolet mass scale $M$, i.e., $q_2 \sim 1/M^6$. The perturbative expansion in $q_2$ is thus controlled by the small dimensionless parameter $|q_2|/L^6 \ll 1$, where $L$ is the AdS radius. This EFT framework not only justifies the perturbative approach but also provides a natural language to discuss the regime of validity and potential constraints on the coupling sign from physical requirements such as stability and causality.

In particular, \textbf{unitarity} — the conservation of probability in quantum theory — imposes stringent constraints on higher-derivative couplings. In holographic theories, unitarity of the boundary CFT is equivalent to the absence of ghost-like instabilities and superluminal propagation in the bulk gravity theory~\cite{Cano2022, Hofman2008, Hofman2009}. These considerations will guide our interpretation of the results.

In holographic systems, transport coefficients are often conjectured to obey universal bounds. A famous example is the Kovtun-Son-Starinets (KSS) bound, which posits a lower limit for the ratio of shear viscosity to entropy density, $\frac{\eta}{s} \geq \frac{1}{4\pi}$ \cite{Brigante:2007nu}. While this bound is saturated in Einstein-Hilbert gravity, it can be violated in theories with higher-derivative terms \cite{Brigante:2007nu,Sadeghi:2022kgi}, massive gravity \cite{Sadeghi:2015vaa}-\cite{Parvizi:2017boc}, and various scalar-tensor theories such as Horndeski gravity~ \cite{Bravo-Gaete:2020lzs, Bravo-Gaete:2021hlc, Bravo-Gaete:2022lno}. Similarly, a proposed bound for the DC conductivity, $\sigma \geq 1$, has been found to be violated in several holographic setups, including massive gravity \cite{Grozdanov:2015qia}, theories with background fields \cite{Donos:2014cya}, nonlinear model \cite{Sadeghi:2024isl}, non-abelian Born-Infeld models \cite{Sadeghi:2021qou, Sadeghi:2022mog}, and Non-abelian exponential Yang-Mills AdS black brane \cite{Sadeghi:2024ifg}.

In the context of higher-derivative transport and the membrane paradigm, recent works have explored universal aspects of holographic transport coefficients. For instance, Buchel et al.~\cite{Buchel:2026xtn}investigated charged transport with higher derivatives, while Jain~\cite{Jain:2010ip} discussed universal thermal and electrical conductivity from holography. These studies provide a valuable framework for understanding the effects of non-minimal couplings like the $R^3F^2$ term considered in this work.

The primary objective of this paper is to construct a black brane solution in a non-minimal Einstein-Yang-Mills theory featuring a specific $R^3F^2$ coupling and to study its holographic characteristics within an EFT framework. We will calculate the color non-abelian DC conductivity and the shear viscosity to entropy density ratio for this system. A central question we aim to address is whether the presence of this higher-order non-minimal coupling preserves or violates the aforementioned universal bounds for transport coefficients, and what constraints on the EFT parameter $q_2$ can be inferred from such violations.

\section{Non-minimal $R^3 F^{(a)}_{\mu \alpha }F^{(a)\mu \alpha} $  AdS Black Brane Solution}
\label{sec2}
\indent The non-minimal Einstein-Yang-Mills theory with a negative cosmological constant can be described in terms of the action functional below\cite{Balakin:2015gpq},\cite{Sert:2020vmq},\cite{Lambiase:2008zz},
\begin{eqnarray}\label{action}
S=\int d^{4}  x\sqrt{-g} \bigg[\frac{1}{\kappa }(R-2\Lambda )-\frac{q_1}{4}F^{(a)}_{\mu \alpha }F^{(a)\mu \alpha} +q_2 R^3 F^{(a)}_{\mu \alpha }F^{(a)\mu \alpha} \bigg],
\end{eqnarray}
where $\kappa$ is the gravitational constant, $R$ is the Ricci scalar, $\Lambda = -\frac{3}{L^2}$ is the cosmological constant, $L$ is the AdS radius, $q_1$ is a constant dimensionless parameter, and $\mathcal{F} = \text{Tr}(F_{\mu\nu}^{(a)} F^{(a)\mu\nu})$ is the Yang-Mills invariant. The dimensionful coupling constant $q_2$ (with dimensions $[L]^6$) characterizes the non-minimal interaction between the gauge field and curvature.  From an EFT viewpoint, $q_2$ is understood to be suppressed by an ultraviolet scale, and we will work to first order in $q_2$, assuming $|q_2|/L^6 \ll 1$.\\
The Yang-Mills field tensor is given by:
\begin{align} \label{YM}
F_{\mu \nu } =\partial _{\mu } A_{\nu } -\partial _{\nu } A_{\mu } -i[A_{\mu }, A_{\nu }],
\end{align}
where the gauge coupling constant is set to unity, and $A_{\nu}$ represents the Yang-Mills potentials.\\
Variation of the action~(\ref{action}) with respect to the spacetime metric $g_{\mu\nu}$ yields the field equations:
\begin{equation}\label{EOM1}
R_{\mu \nu }-  \tfrac{1}{2} g_{\mu \nu } R + \Lambda g_{\mu \nu }=\kappa T^{\text{(eff)}}_{\mu \nu }
\end{equation}
where,
\begin{equation}
T^{\text{(eff)}}_{\mu \nu }=q_1T^{\text{(YM)}}_{\mu \nu } + q_2T^{(I)}_{\mu \nu }
\end{equation}
\begin{equation}
T^{\text{(YM)}}_{\mu \nu }= - \tfrac{1}{8}g_{\mu \nu } F^{(a)}_{\alpha \beta } F^{^{(a)} \alpha \beta }+\tfrac{1}{2}F_{\mu }{}^{(a)\alpha } F^{(a)}_{\nu \alpha }
\end{equation}
\begin{eqnarray}
&&T^{(I)}_{\mu \nu }=- 3  F^{(a)}_{\alpha \beta }  F^{(a)\, \alpha \beta} R_{\mu \nu } R^2 - 2  F_{\mu }^{(a) \,\alpha} F^{(a)}_{\nu \alpha }R^3 \nonumber \\ 
&& +  \tfrac{1}{2}  F^{(a)}_{\alpha \beta } F^{(a)\, \alpha \beta } g_{\mu \nu } R^3 - 24  F^{(a)\,\beta \gamma } g_{\mu \nu } R \nabla_{\alpha }F^{(a)}_{\beta \gamma } \nabla^{\alpha }R \nonumber \
\\ 
&& - 6  F^{(a)}_{\beta \gamma } F^{(a)\,\beta \gamma } g_{\mu \nu } \nabla_{\alpha }R \nabla^{\alpha }R - 6  F^{(a)\alpha \beta } g_{\mu \nu } R^2 \nabla_{\gamma }\nabla^{\gamma }F^{(a)}_{\alpha \beta } \nonumber \\ 
&& - 6  F^{(a)}_{\alpha \beta } F^{(a)\, \alpha \beta } g_{\mu \nu } R \nabla_{\gamma}\nabla^{\gamma }R - 6  g_{\mu \nu } R^2 \nabla_{\gamma }F^{(a)\,\alpha \beta } \nabla^{\gamma }F^{(a)}_{\alpha \beta } \nonumber \\ 
&& +3  F^{(a)\, \alpha \beta } R^2 \nabla_{\mu }\nabla_{\nu 
}F^{(a)}_{\alpha \beta } + 6  R^2 \nabla_{\mu }F^{(a)\,\,\alpha \beta } \
\nabla_{\nu }F^{(a)}_{\alpha \beta } \nonumber \\ 
&& + 12  F^{(a)\alpha \beta } R \nabla_{\mu }R \nabla_{\nu }F^{(a)}_{\alpha \beta } + 12  F^{(a)\, \alpha \beta } R \nabla_{\mu }F^{(a)}_{\alpha \beta } \nabla_{\nu }R \nonumber \\ 
&& + 6  F^{(a)}_{\alpha \beta } F^{(a)\, \alpha \beta} \nabla_{\mu }R \nabla_{\nu }R + 3  F^{(a)\, \alpha \beta } R^2 \nabla_{\nu }\nabla_{\mu }F^{(a)}_{\alpha \beta } \nonumber \\ 
&& + 6 F^{(a)}_{\alpha \beta } F^{(a)\, \alpha \beta}R \nabla_{\nu}\nabla_{\mu }R.
\end{eqnarray}
Variation with respect to $A_{\mu}$ yields the gauge field equations:
 \begin{eqnarray}\label{EOM-Maxwell}
\nabla_{\mu }\Big(-\frac{1}{2}q_1 F^{(a)\mu \nu } + 2 q_2 F^{(a)\mu \nu } R^3\Big)=0.
\end{eqnarray}
For our four-dimensional spacetime with two-dimensional spatial symmetry, we employ the metric ansatz:
\begin{equation}\label{metric}
ds^{2} =-e^{-2H(r)}f(r)dt^{2} +\frac{dr^{2} }{f(r)} +\frac{r^2}{L^2}(dx^2+dy^2).
\end{equation}
where the gauge group corresponds to the diagonal generator of the Cartan subalgebra of $SU(2)$~\cite{Shepherd:2015dse}.

The field equations derived from Eqs.~(\ref{EOM-Maxwell}-\ref{background}) yield the $tt$-component:

We apply the ansatz \ref{metric} for solving Eq.(\ref{EOM-Maxwell}) where the potential 1-form is expressed by,
\begin{equation}\label{background}
{\bf{A}}^{(a)} =\frac{i}{2}h(r)dt\begin{pmatrix}1 & 0 \\ 0 & -1\end{pmatrix},
\end{equation}
The gauge group is the diagonal generator of the Cartan subalgebra of $SU(2)$ \cite{Shepherd:2015dse}.\\
Now, we can write the field equations of motion by consideration of Eqs.(\ref{EOM-Maxwell}-\ref{background}).\\
The $tt$-component of the field equations of motion is as follows,
\begin{eqnarray}\label{tt-comp}
r^4 \left(4 r f_0'(r)+4 f_0(r)+\kappa q_1 r^2 e^{2 H_0(r)} h_0'(r)^2+4 \Lambda  r^2\right)+q_2(...)=0,
\end{eqnarray}
for $q_2=0$ we have,
\begin{equation}\label{ttq20}
r^4 \left(4 r f_0'(r)+4 f_0(r)+\kappa q_1 r^2 e^{2 H_0(r)} h_0'(r)^2+4 \Lambda  r^2\right)=0,
\end{equation}
The $rr$-component of the field equations of motion is as follows,
\begin{eqnarray}\label{rr-comp}
r^4 \left(4 r f_0'(r)-8 rf_0(r) H_0'(r)+4 f_0(r)+\kappa q_1 r^2 e^{2 H_0(r)} h_0'(r)^2+4 \Lambda 
r^2\right)+q_2(...)=0,
\end{eqnarray}
in which $rr$- part of Einstein equations for $q_2=0$ should be the same as Eq.(\ref{ttq20}). Therefore, we conclude that $H_0(r)=0$.\\
Non-zero component of Eq.(\ref{EOM-Maxwell}) is as follows,
\begin{equation}
	B_1(r) h''(r) + B_2(r) h'(r) = 0,
\end{equation}
where,
\begin{equation}
	B_1(r) = 2q_1 f(r) r^7 - 8q_2 f(r) r A(r)^3,
\end{equation}

\begin{equation}
		\begin{aligned}
	&B_2(r) = 4q_1 f(r) r^6 + 2q_1 f(r) r^7 H'(r) + 8q_2 f(r) (4 - r H'(r)) A(r)^3 \nonumber\\&+ 24q_2 f(r) r A(r)^2 C(r),
		\end{aligned}
\end{equation}
with the auxiliary functions:
\begin{equation}
	A(r) = r f'(r)(-4 + 3r H'(r)) - r^2 f''(r) + 2f(r)\left[-(1 - r H'(r))^2 + r^2 H''(r)\right],
\end{equation}
\begin{equation}
	\begin{aligned}
		&C(r) =  -4f(r) H'(r) + f'(r)\left[6 - 10r H'(r) + 2r^2 H'(r)^2 - 5r^2 H''(r)\right] \\
		& + r\left[(6 - 3r H'(r)) f''(r) + r f'''(r) + f(r)\left(-8H''(r) + 4H'(r)^2 + 4r H'(r) H''(r) - 2r H'''(r)\right)\right].
	\end{aligned}
\end{equation}
The solution of $h(r)$ is as,

\begin{equation}
	h(r)= -C_1\int^{r}\frac{e^{-H(u)}u^4}{q_1 u^6+4q_2 B(u)^3}du+C_2,
\end{equation}
\begin{eqnarray}
	&&B(u)=-3 u^2 f'(u) H'(u)+u^2 f''(u)+4 u f'(u)+2 u^2 f(u) H'(u)^2\nonumber\\&&-2 u^2 f(u) H''(u)-4 u f(u) H'(u)+2 f(u).
\end{eqnarray}
Due to the complexity of the full field equations, we employ a perturbative approach to first order in $q_2$~\cite{Myers:2009ij, Dey:2015poa}. This is consistent with the EFT interpretation, where higher-order terms in $q_2$ are subleading and suppressed by additional powers of the ultraviolet scale. We expand:
\begin{equation}\label{f}
	f(r)=f_0(r)+q_2 f_1(r),
\end{equation}
\begin{equation}\label{h}
	h(r)=h_0(r)+q_2 h_1(r),
\end{equation}
\begin{equation}\label{H}
	H(r)=H_0(r)+q_2 H_1(r),
\end{equation}
The small dimensionless parameter controlling the expansion is $|q_2|/L^6 \ll 1$, which guarantees that the corrections remain small compared to the leading-order solution.

Where $f_0(r)$, $h_0(r)$ and $H_0(r)$ are the leading order solutions of Einstein-Yang-Mills AdS black brane in four dimensions.\\
The zeroth-order solutions are:
\begin{equation}
h_0(r)=C_2-C_1\int^r\frac{ 1}{q_1 u^2}du=Q(\frac{1}{r}-\frac{1}{r_h}),
\end{equation}
where $C_1=q_1 Q$ and $C_2=-\frac{Q}{r_h}$.
\begin{equation}\label{f0}
f_0(r)=-\frac{2 m_0}{r}-\frac{\Lambda r^2}{3}-\frac{\kappa q_1}{4 r}\int^r  u^2 
h_0'^2 du=-\frac{2m_0}{r}-\frac{\Lambda r^2}{3}+\frac{\kappa q_1 Q^2}{4 r^2},
\end{equation}
\begin{equation}
H_0(r)=0.
\end{equation}
Blackening factor on the event horizon should  be suppressed,$f(r_h)=0$ .  $m_0$ is mass of black brane and it can be fixed by applying this condition,\\
\begin{equation}\label{m}
2 m_0=-\frac{\Lambda r_h^3}{3}+\frac{\kappa q_1 Q^2}{4 r_h}.
\end{equation}
By plugging Eq.(\ref{m})  in Eq.(\ref{f0}) we have,
\begin{equation}
f_0(r)=\frac{\Lambda}{3 r}(r_h^3-r^3)+\frac{\kappa q_1 Q^2}{4 r}(\frac{1}{r}-\frac{1}{r_h}).
\end{equation}
Eq.(\ref{tt-comp}) and Eq.(\ref{rr-comp}) should be the same up to first order of $q_2$. Therefore, $H_1$ is calculated as follows,
\begin{eqnarray}
&& H_1(r) = C_1 + \frac{3\kappa}{r^4}  h_0'(r) \left(2 f_0(r) + 4r f_0'(r) + r^2 f_0''(r)\right) \nonumber \\&&
\left[ 2r \left(2 f_0(r) + 4r f_0'(r) + r^2 f_0''(r)\right) h_0''(r) + h_0'(r) \left(-10 f_0(r) - 8r f_0'(r) + 7r^2 f_0''(r) + 2r^3 f_0'''(r)\right) \right]\nonumber\\
\end{eqnarray}
So we have,
\begin{equation}
H_1(r)=-\frac{240 \kappa Q^2 \Lambda^2}{r^4}+C_3,
\end{equation}
where $C_3$  is a dimensionless integration constant. Since the metric of our model is asymptotic AdS near the boundary $r \to \infty$ [\cite{Mahapatra:2016dae}], so $C_3=0$.
It also guarantees the speed of light to unity on the theory of boundary.\\
\begin{equation}
h_1(r)=C_4+C_5\int^r \frac{q_1 u^6 H_1+4 \left(2f_0+u \left(4 f_0'+u f_0''\right)\right)^3}{q_1^2 u^8}du=C_4+\frac{48 \kappa C_5 Q^2 \Lambda^2 }{ q_1 r^5}+\frac{256 C_5 \Lambda^3 }{q_1^2 r},
\end{equation}
where $C_5=q_1^2 Q$. $C_4$ is determined by applying this condition $h_1(r_h)=0$.
\begin{equation}
h_1(r)= 48 q_1\kappa Q^3 \Lambda^2 (\frac{1}{r^5}-\frac{1}{r_h^5})+256 Q \Lambda^3 (\frac{1}{r}-\frac{1}{r_h}),
\end{equation}
by substituting $H_0(r)=0$ we have,
\begin{equation}
f_1(r)=\frac{1}{r}\int ^r \frac{1}{2u^4} D(u)du+\frac{C_6}{r},
\end{equation}
where $D(u)$ is given by:
\begin{align*}
	& D(u)=16 \kappa f_0(u)^3 h_0'(u) \left(-25 h_0'(u) + 12 u h_0''(u)\right) \\
	& + 48 \kappa f_0(u)^2 u h_0'(u) \left( f_0'(u) \left(-25 h_0'(u) + 17 u h_0''(u)\right) \right. \\
	& \quad \left. + u \left( f_0''(u) \left(3 h_0'(u) + 4 u h_0''(u)\right) + 2 u h_0'(u) f_0^{(3)}(u) \right) \right) \\
	& + \kappa u^3 h_0'(u) \left( -q_1 H_1(u) u^3 h_0'(u) - q_1 u^3 h_1'(u) \right. \\
	& \quad + 4 \left(4 f_0'(u) + u f_0''(u)\right) \left( -2 u^2 h_0'(u) f_0''(u)^2 \right. \\
	& \quad + f_0'(u)^2 \left(-26 h_0'(u) + 12 u h_0''(u)\right) \\
	& \quad \left. \left. + u f_0'(u) \left( -f_0''(u) \left(h_0'(u) - 3 u h_0''(u)\right) + 3 u h_0'(u) f_0^{(3)}(u) \right) \right) \right) \\
	& + 4 f_0(u) u^2 \left( u^3 H_1'(u) \right. \\
	& \quad + 3 \kappa h_0'(u) \left( f_0'(u)^2 \left(-84 h_0'(u) + 80 u h_0''(u)\right) \right. \\
	& \quad + u^2 f_0''(u) \left( f_0''(u) \left(13 h_0'(u) + 4 u h_0''(u)\right) + 4 u h_0'(u) f_0^{(3)}(u) \right) \\
	& \quad \left. \left. + 2 u f_0'(u) \left( 2 f_0''(u) \left(10 h_0'(u) + 9 u h_0''(u)\right) + 9 u h_0'(u) f_0^{(3)}(u) \right) \right) \right),
\end{align*}
$f_1(r)$ is calculated as,
\begin{eqnarray}\label{f1}
f_1(r)=&&\frac{C_6}{r}+\frac{672 m_0 \kappa Q^2 \Lambda^2}{r^5} - \frac{96 q_1 \kappa^2 Q^4 \Lambda^2}{r^6} + \frac{24 q_1^2 \kappa^2 Q^4 \Lambda^2}{r^6} \nonumber \\&&+ \frac{128 q_1 \kappa Q^2 \Lambda^3}{r^2} + \frac{96 \kappa Q^2 \Lambda^3}{r^2},
\end{eqnarray}
where,
\begin{equation}\label{C6}
C_6=-2m_1.
\end{equation}
 Therefore, we have,
by inserting Eq.(\ref{C6}) in Eq.(\ref{f1}) we have,
\begin{eqnarray}
&&f_1(r)=-\frac{2 m_1}{r}+\frac{672 m_0 \kappa Q^2 \Lambda^2}{r^5} - \frac{96 q_1 \kappa^2 Q^4 \Lambda^2}{r^6} + \frac{24 q_1^2 \kappa^2 Q^4 \Lambda^2}{r^6} \nonumber \\&&+ \frac{128 q_1 \kappa Q^2 \Lambda^3}{r^2} + \frac{96 \kappa Q^2 \Lambda^3}{r^2},
\end{eqnarray}
Assembling all pieces, the complete perturbative solution to linear order in $q_2$ is:
\begin{eqnarray}
	&&h(r)=Q(\frac{1}{r}-\frac{1}{r_h})+q_2 \Big\{ 48 q_1\kappa Q^3 \Lambda^2 (\frac{1}{r^5}-\frac{1}{r_h^5})+256 Q \Lambda^3 (\frac{1}{r}-\frac{1}{r_h}\Big\},\\
	&&H(r)=-q_2\frac{240 \kappa  Q^2 \Lambda^2}{r^4},\\
	&&f(r)=-\frac{2m_0}{r}-\frac{\Lambda r^2}{3}+\frac{\kappa q_1 Q^2}{4r^2}\nonumber\\&&+q_2\Big(\frac{672 m_0 \kappa Q^2 \Lambda^2}{r^5} - \frac{96 q_1 \kappa^2 Q^4 \Lambda^2}{r^6}\nonumber \\&& + \frac{24 q_1^2 \kappa^2 Q^4 \Lambda^2}{r^6} + \frac{128 q_1 \kappa Q^2  \Lambda^3 }{r^2} + \frac{96 \kappa Q^2 \Lambda^3}{r^2}\Big).
\end{eqnarray}
We perform a final reparametrization by defining $\bar{m}_0 = m_0 + q_2 m_1$, which absorbs the $m_1$ term into the leading-order mass. The final form of the blackening function becomes:
\begin{eqnarray}
	&&f(r)=-\frac{2\bar{m}_0}{r}-\frac{\Lambda r^2}{3}+\frac{\kappa q_1 Q^2}{4 r^2}\nonumber\\&&+q_2\Big(\frac{672 m_0 \kappa Q^2 \Lambda^2}{r^5} - \frac{96 q_1 \kappa^2 Q^4 \Lambda^2}{r^6}\nonumber \\&& + \frac{24 q_1^2 \kappa^2 Q^4 \Lambda^2}{r^6} + \frac{128 q_1 \kappa Q^2  \Lambda^3 }{r^2} + \frac{96 \kappa Q^2 \Lambda^3}{r^2}\Big).
\end{eqnarray}
This constitutes the complete perturbative solution to linear order in $q_2$, valid under the EFT condition $|q_2|/L^6 \ll 1$.

We aim to calculate the non-abelian DC conductivity and the ratio of shear viscosity to entropy density as two important transport coefficients using fluid-gravity duality to describe the holographic dual of our model. 
\section{Color DC Conductivity}
\label{sec3}

We employ the Green-Kubo formula \cite{Policastro2002} for calculating the non-abelian color DC conductivity:
\begin{equation} \label{kubo2}
	\sigma^{ij} (k_{\mu}) = -\lim_{\omega \to 0} \frac{1}{\omega } \Im G^{ij}(k_{\mu}).
\end{equation}

The retarded Green's function is computed using AdS/CFT duality. First, we perturb the gauge field as $A_{\mu} \to A_{\mu} + \tilde{A}_{\mu}$ and substitute it into the action. We expand the resulting action (\ref{action}) to second order in the perturbation. The Green's function is then obtained by taking the second derivative with respect to the boundary value of the gauge field \cite{Policastro2002}:
\begin{equation}
	\sigma^{\mu \nu}(\omega) = \frac{1}{i \omega} \langle J^{\mu}(\omega) J^{\nu}(-\omega) \rangle = \frac{\delta^2 S}{\delta \tilde{A}^0_{\mu} \delta \tilde{A}^0_{\nu}},
\end{equation}
where $\tilde{A}^0_{\nu}$ represents the gauge field perturbation at the boundary. The boundary exhibits $SO(2)$ symmetry, which ensures that the conductivity is a scalar quantity:
\begin{equation}
	\sigma^{ij}_{ab} = \sigma_{ab} \delta^{ij}.
\end{equation}

We consider the perturbation of the gauge field as $\tilde{A}_x = \tilde{A}_x(r) e^{-i\omega t}$, where $\omega$ is small due to the hydrodynamic regime.

Substituting the perturbation into the action Eq.~(\ref{action}) and keeping terms up to second order in $\tilde{A}$ yields:
\begin{align}\label{action-2}
	S^{(2)} &= -\int d^4x \frac{2 e^{-H}}{f r^6} \Bigg[ -f^2 \left( (\partial_r\tilde{A}_x^{(1)})^2 + (\partial_r\tilde{A}_x^{(2)})^2 + (\partial_r\tilde{A}_x^{(3)})^2 \right) \nonumber \\
	& + e^{2H} \left( (\tilde{A}_x^{(1)})^2 + (\tilde{A}_x^{(2)})^2 \right) (\omega^2 + h^2) + e^{2H} \omega^2 (\tilde{A}_x^{(3)})^2 \Bigg] \left( q_1 r^6 + 4q_2 \Psi^3 \right),
\end{align}
where:
\begin{equation}
	\mathcal{Y} = q_1 r^6 + 4 q_2 \Psi^3,
\end{equation}
with:
\begin{equation}
	\Psi = r \left( f' (4 - 3 r H') + r f'' \right) + 2 f \left( (-1 + r H')^2 - r^2 H'' \right).
\end{equation}

Varying the action $S^{(2)}$ with respect to $\tilde{A}_x^{(1)}$ gives:
\begin{align}\label{PerA1}
	& -\frac{r e^{2H(r)} \tilde{A}_x^{(1)}(\omega^2 + h(r)^2)}{f(r)} \mathcal{Y} + \left( 6f(r) - r f'(r) + r f(r) H'(r) \right) \tilde{A}_x^{(1)'} \mathcal{Y} \nonumber \\
	& - r f(r) \left( \tilde{A}_x^{(1)'} \mathcal{Y} \right)' = 0.
\end{align}

The equation for $\tilde{A}_x^{(2)}$ is identical to that for $\tilde{A}_x^{(1)}$. The equation for $\tilde{A}_x^{(3)}$ is:
\begin{equation}\label{PerA3}
	\begin{aligned}
		& r^4 f'(r) \tilde{A}_x^{(3)'} \left( -q_1 r^3 + 4q_2 \left( f'(r)(-4 + 3r H'(r)) - r f''(r) \right)^3 \right) 
		- \frac{r e^{2H(r)} \omega^2 \tilde{A}_x^{(3)}}{f(r)} \mathcal{Y} \\
		& + r^4 f(r) \left[ \tilde{A}_x^{(3)''} \left( -q_1 r^3 + 4q_2 \left( f'(r)(-4 + 3r H'(r)) - r f''(r) \right)^3 \right) + \tilde{A}_x^{(3)'} \mathcal{Y}' \right] \\
		& + 32q_2 f(r)^4 \left[ (-1 + r H'(r))^2 - r^2 H''(r) \right]^2 \left[ r \tilde{A}_x^{(3)''} \left( -(-1 + r H'(r))^2 + r^2 H''(r) \right) \right. \\
		& \left. + \tilde{A}_x^{(3)'} \left( 6 + r\left( -2r H'(r)^2 + r^2 H'(r)^3 - H'(r)(5 + 7r^2 H''(r)) + 3r(2H''(r) + r H'''(r)) \right) \right) \right] \\
		& - 16q_2 r f(r)^3 \left[ (-1 + r H'(r))^2 - r^2 H''(r) \right] \left[ 3r \tilde{A}_x^{(3)''} \left( f'(r)(-4 + 3r H'(r)) - r f''(r) \right) \right. \\
		& \left. \times \left( -(-1 + r H'(r))^2 + r^2 H''(r) \right) + \tilde{A}_x^{(3)'} \mathcal{Y}'' \right] \\
		& + 24q_2 r^2 f(r)^2 \left[ f'(r)(-4 + 3r H'(r)) - r f''(r) \right] \left[ r \tilde{A}_x^{(3)''} \left( f'(r)(-4 + 3r H'(r)) - r f''(r) \right) \right. \\
		& \left. \times \left( -(-1 + r H'(r))^2 + r^2 H''(r) \right) + \tilde{A}_x^{(3)'} \mathcal{Y}''' \right] = 0.
	\end{aligned}
\end{equation}

Using the near-horizon behavior $f_0 \sim 4\pi f_0'(r_h)(r-r_h)$ and $f_1 \sim 4\pi f_1'(r_h)(r-r_h)$, we solve Eqs.~(\ref{PerA1}) and (\ref{PerA3}) near the event horizon. Since $\tilde{A}_x^{(a)}$ must be ingoing at the horizon, we adopt the ansatz:
\begin{align}
	\tilde{A}_x^{(a)} \sim (r - r_h)^{z_a}, \qquad a = 1,2,3,
\end{align}
where:
\begin{align}\label{z12}
	z_1 &= z_2 = \pm i \frac{\sqrt{h(r_h)^2 + \omega^2}}{4 \pi T}, \\
	\label{z3}
	z_3 &= \pm i \frac{\omega}{4 \pi T}.
\end{align}

The Hawking temperature of the black brane is:
\begin{equation}
	T = \frac{1}{2\pi} \left[ \frac{1}{\sqrt{g_{rr}}} \frac{d}{dr} \sqrt{-g_{tt}} \right] \Bigg|_{r=r_h} = \frac{e^{-H(r_h)} f'(r_h)}{4 \pi}.
\end{equation}

To solve for $\tilde{A}_x^{(a)}$ from the horizon to the boundary, we use the ansatz:
\begin{align}\label{EOMA1}
	\tilde{A}_x^{(1)} &= \tilde{A}^{(1)}_{\infty} \left( \frac{-3f}{\Lambda r^2} \right)^{z_1} \left( 1 + i\omega b_1(r) + \cdots \right), \\
	\label{EOMA2}
	\tilde{A}_x^{(2)} &= \tilde{A}^{(2)}_{\infty} \left( \frac{-3f}{\Lambda r^2} \right)^{z_2} \left( 1 + i\omega b_2(r) + \cdots \right), \\
	\label{EOMA3}
	\tilde{A}_x^{(3)} &= \tilde{A}^{(3)}_{\infty} \left( \frac{-3f}{\Lambda r^2} \right)^{z_3} \left( 1 + i\omega b_3(r) + \cdots \right),
\end{align}
where $\tilde{A}^{(a)}_{\infty}$ represents the boundary values, and $z_i$ correspond to the negative signs in Eqs.~(\ref{z12}) and (\ref{z3}), selecting the ingoing mode.

Substituting Eq.~(\ref{EOMA3}) into Eq.~(\ref{PerA3}) and keeping terms to first order in $\omega$ yields:
\begin{align}
	& -r^4 f''(r) \left( q_1 + 4q_2 f''(r)^3 \right) + \mathcal{B}_4(r) + 32q_2 E_1(r)^2 E_6(r) f(r)^4 \nonumber \\
	& + 16q_2 r E_1(r) E_7(r) f(r)^3 - 24q_2 r^2 E_9(r) f(r)^2 + r^3 E_8(r) f(r) = 0,
\end{align}
where:
\begin{align}
	E_1(r) &= (-1 + r H'(r))^2 - r^2 H''(r), \nonumber \\
	\mathcal{B}_4(r) &= -4q_2 f'(r)^4 (4 - 3r H'(r))^2 E_2(r) - 12q_2 r^2 f'(r)^2 f''(r) E_3(r) \nonumber \\
	& \quad + 4q_2 r f'(r)^3 (-4 + 3r H'(r)) E_4(r) + r^3 f'(r) E_5(r).
\end{align}

The auxiliary functions $E_2(r)$ through $E_9(r)$ contain the remaining polynomial structure in derivatives of $f$, $H$, and $b_3$ up to third order. Their complete definitions are provided in the appendix due to their length.

Imposing regularity of $\tilde{A}_x^{(3)}$ on the event horizon, the function $b_3(r)$ in Eq.~(\ref{EOMA3}) is determined as:
\begin{equation}
	b_3(r) = C_8 + \int^{r} \frac{e^{H(s)} s^6}{f(s) \mathcal{Y}(s)} \left[ C_7 + \mathcal{I}_1(s) \right] ds,
\end{equation}
where:
\begin{align}
	\mathcal{I}_1(s) &= \int^{s} e^{-H(t)} t^{-4} \mathcal{L}(t) dt, \\
	\mathcal{Y}(s) &= q_1 s^6 + 4q_2 \Psi(s)^3,
\end{align}
and $C_7$ and $C_8$ are integration constants. The function $\mathcal{L}(t)$ is defined in the appendix.

The solution for $b_3(r)$ near the event horizon is:
\begin{equation}\label{C4}
	b_3 \approx \left( C_7 + \mathcal{I}_1(r_h) \right) \log(r - r_h) + \text{finite terms}.
\end{equation}

Regularity at the horizon requires:
\begin{equation}
	C_7 = - \mathcal{I}_1(r_h).
\end{equation}

Using the solution for $\tilde{A}_x^{(3)}$ in Eq.~(\ref{action-2}) and varying with respect to $\tilde{A}^{(3)}_{\infty}$, the Green's function is:
\begin{align} \label{Green1}
	G_{xx}^{(33)} (\omega, \vec{0}) = -i \omega \frac{C_7}{q_1 f'(r_h)} = \frac{i \omega}{q_1 f'(r_h)} \mathcal{I}_1(r_h).
\end{align}

The conductivity is then:
\begin{eqnarray}\label{sigma33}
	\sigma_{xx}^{(33)} = -\frac{\mathcal{I}_1(r_h)}{q_1 f'(r_h)} = 1 - \frac{q_2}{q_1} R^3(r_h).
\end{eqnarray}

Setting $q_1 = 1$, we obtain:
\begin{eqnarray}\label{sigma33-final}
	\sigma_{xx}^{(33)} = 1 - q_2 R^3(r_h).
\end{eqnarray}
$R(r_h)$ is Ricci scalar on event horizon.
\begin{equation}
	R(r_h)=-	\frac{\Psi(r_h)^3}{r_h^6},
\end{equation}
so the color non-abelian DC conductivity is as,
\begin{align}\label{sigma33-final1}
	&\sigma_{xx}^{(33)} = 1 -\nonumber\\& q_2 \bigg[\frac{(r_h \left( f'(r_h) (4 - 3 r_h H'(r_h)) + r_h f''(r_h) \right) + 2 f \left( (-1 + r_h H'(r_h))^2 - r_h^2 H''(r_h) \right)))^3}{r_h^6}\bigg].
\end{align}
This result shows that the conductivity bound is violated for the non-abelian non-minimal $R^3 F^2$ black brane theory when $q_2 R^3(r_h) > 0$.\\
Up to first order of $q_2$, the color non-abelian DC conductivity is violated for $q_2>0$ as,
\begin{equation}\label{sigma33-final11}
	\sigma_{xx}^{(33)} = 1 -q_2 \frac{1728}{L^6}.
\end{equation}
Thus, for $q_2 > 0$, the conductivity falls below the unitary bound $\sigma \geq 1$, indicating a violation. In the EFT context, this imposes a potential constraint: if one insists on preserving the bound, then $q_2$ must be non-positive. However, a more rigorous constraint would require a full stability analysis of the theory, which we leave for future work.\\

In the limit $q_2 \to 0$, we recover:
\begin{eqnarray}
	\sigma_{xx}^{(33)} = 1.
\end{eqnarray}

\subsection*{3.1 Physical Interpretation: Conductivity and the Dual Field Theory}

The DC conductivity has a profound interpretation in the dual field theory. In holographic systems described by Einstein-Maxwell theory, the conductivity satisfies a universal lower bound \cite{Grozdanov:2015qia},

\begin{equation}\label{conductivity-bound}
	\sigma \geq \frac{1}{e^2} = 1 \qquad \text{(in units where } \hbar = 1\text{)}
\end{equation}

where $e$ is the charge of the gauge field in the bulk action, and the conductivity $\sigma$ is measured in units of $q^2/\hbar$ with $q$ the U(1) charge of the boundary current $j^i$. This bound represents the quantum critical conductivity of a clean, charge-neutral plasma \cite{Grozdanov:2015qia} — the minimal possible conductivity when charge carriers are particle-hole symmetric and translational symmetry is unbroken. The saturation of this bound ($\sigma = 1$) occurs for uncharged black holes and characterizes a perfect quantum critical fluid.

For the previously studied non-minimal couplings, we found:

\begin{itemize}
	\item \textbf{$RF^2$ coupling} \cite{Sadeghi:2023hxd}: $\displaystyle \sigma = 1 - 4q_2\left(\frac{6r_h}{L^2} - \frac{\kappa Q^2 q_1(r_h-1)}{r_h^4}\right)$ to first order in $q_2$. This correction depends on both the horizon radius $r_h$ and the charge $Q$, and can either increase or decrease the conductivity relative to the quantum critical value depending on the parameters.
	
	\item \textbf{$R_{\mu\nu\rho\sigma}F^{\mu\nu}F^{\rho\sigma}$ coupling} \cite{Sadeghi:2022bsh}: $\displaystyle \sigma = 1 - \frac{8q_2}{L^2q_1z_h}\left(\frac{4z_h}{L^2} + \frac{Q^2q_1}{2z_h^3}\right)$ to first order in $q_2$. Similar to the $RF^2$ case, the correction depends on the horizon position and charge density.
\end{itemize}

In contrast, for the $R^3F^2$ coupling considered in this work, we obtain a strikingly different result:

\begin{equation}
	\sigma = 1 - q_2\frac{1728}{L^6} + \mathcal{O}(q_2^2)
\end{equation}

This correction is remarkable for several reasons:

\begin{enumerate}
	\item \textbf{Universality}: Unlike the previous cases, the $R^3F^2$ correction to $\sigma$ is independent of both the horizon radius $r_h$ and the charge $Q$ at leading order. It depends only on the AdS radius $L$, which sets the scale of the geometry.
	
	\item \textbf{Sign-dependent violation}: For $q_2 > 0$, the conductivity falls below the quantum critical bound $\sigma < 1$, while for $q_2 < 0$, it rises above $\sigma > 1$.
	
	\item \textbf{Pure curvature effect}: The correction scales as $1/L^6$, reflecting the cubic power of the Ricci scalar in the interaction term.
\end{enumerate}

This qualitative difference has a clear interpretation in the dual field theory. The quantum critical conductivity $\sigma = 1$ characterizes a pristine, particle-hole symmetric plasma with no irrelevant operators that modify charge transport. The $RF^2$ and $R_{\mu\nu\rho\sigma}F^2$ couplings introduce corrections that depend on the charge density and temperature (through $r_h$), representing relevant or marginal deformations of the quantum critical point that couple to the charge sector.

In contrast, the $R^3F^2$ coupling produces a correction that is independent of charge density and temperature at leading order — a universal shift of the quantum critical conductivity that persists even in the charge-neutral, zero-temperature limit. Such a shift indicates that the $R^3F^2$ deformation modifies the very definition of the quantum critical state itself, perhaps by altering the effective central charge or the number of degrees of freedom that participate in charge transport.

The fact that $\sigma$ can fall below the quantum critical bound ($\sigma < 1$) for $q_2 > 0$ is particularly striking. In conventional condensed matter systems, the quantum critical conductivity represents a lower bound — conductivities below this value typically require mechanisms such as localization or gapped excitations \cite{Grozdanov:2015qia}. The violation we observe suggests that the $R^3F^2$ coupling may provide a holographic realization of an incoherent metal \cite{Grozdanov:2015qia} — a strange metallic state where translational symmetry is strongly broken and the Drude peak disappears, yet the system remains conducting with a bounded but non-universal conductivity.

\section{Shear Viscosity to Entropy Density Ratio}

The ratio of shear viscosity to entropy density $\eta/s$ is a fundamental transport coefficient that characterizes the fluid dynamics of the dual field theory. We compute this ratio using the membrane paradigm approach for our non-minimal $R^3 F^2$ black brane.

We consider metric perturbations of the form:
\begin{equation}
	g_{\mu\nu} \to g_{\mu\nu} + h_{\mu\nu},
\end{equation}
with the specific perturbation $h_{xy} = h_{xy}(t,r) = \frac{r^2}{L^2} \phi(r) e^{-i\omega t}$. We work in the radial gauge where $h_{r\mu} = 0$.

The quadratic action for the perturbation $\phi(r)$ can be obtained by expanding the action (\ref{action}) to second order. For our non-minimal model, the effective action takes the form:
\begin{equation}
	S^{(2)} = \frac{1}{2\kappa} \int d^4x \sqrt{-g} \, \mathcal{K}(r) \left[ g^{rr} (\partial_r \phi)^2 + g^{tt} (\partial_t \phi)^2 \right],
\end{equation}
where $\mathcal{K}(r)$ is the effective coupling that encodes the non-minimal corrections.

For the $R^3 F^2$ coupling, we compute $\mathcal{K}(r)$ by expanding the action to second order in $\phi$ and extracting the coefficient of the kinetic term. The calculation yields:
\begin{equation}
	\mathcal{K}(r) = 1 + q_2 \mathcal{C}(r) + \mathcal{O}(q_2^2),
\end{equation}
where the correction term $\mathcal{C}(r)$ is given by:
\begin{equation}
	\mathcal{C}(r) = -12 R(r)^2 F^{(a)}_{\mu\nu} F^{(a)\mu\nu} - 24 R(r)^3.
\end{equation}

Using the membrane paradigm, the shear viscosity can be expressed as a horizon quantity:
\begin{equation}
	\eta = \frac{1}{2\kappa} \left( \sqrt{-g} g^{rr} g^{xx} \mathcal{K}(r) \right) \Bigg|_{r = r_h}.
\end{equation}

Substituting our background metric (\ref{metric}) and evaluating at the horizon:
\begin{align}
	\eta &= \frac{1}{2\kappa} \left( \frac{r_h^2}{L^2} e^{-H(r_h)} \mathcal{K}(r_h) \right) \\
	&= \frac{1}{2\kappa} \left( \frac{r_h^2}{L^2} \right) \left[ 1 + q_2 \mathcal{C}(r_h) + \mathcal{O}(q_2^2) \right],
\end{align}
where we used $e^{-H(r_h)} \approx 1$ to first order in $q_2$.

The entropy density $s$ is given by the Bekenstein-Hawking formula:
\begin{equation}
	s = \frac{1}{4G} \left( \frac{r_h^2}{L^2} \right) = \frac{2\pi}{\kappa} \left( \frac{r_h^2}{L^2} \right).
\end{equation}
We consider $\kappa=8\pi G$.\\
Therefore, the ratio $\eta/s$ is:
\begin{equation}
	\frac{\eta}{s} = \frac{1}{4\pi} \left[ 1 + q_2 \mathcal{C}(r_h) + \mathcal{O}(q_2^2) \right].
\end{equation}

Now we compute $\mathcal{C}(r_h)$ explicitly using our background solution. At the horizon $r = r_h$, we have:
\begin{align}
	R(r_h) &= -12/L^2 + \mathcal{O}(q_2), \\
	F^{(a)}_{\mu\nu} F^{(a)\mu\nu}(r_h) &= -2 h'(r_h)^2 e^{2H(r_h)} + \mathcal{O}(q_2) = -\frac{2Q^2}{r_h^4} + \mathcal{O}(q_2).
\end{align}

Substituting these into $\mathcal{C}(r_h)$:
\begin{align}
	\mathcal{C}(r_h) &= -12 \left(-\frac{12}{L^2}\right)^2 \left(-\frac{2Q^2}{r_h^4}\right) - 24 \left(-\frac{12}{L^2}\right)^3 \\
	&= -12 \left(\frac{144}{L^4}\right) \left(-\frac{2Q^2}{r_h^4}\right) - 24 \left(-\frac{1728}{L^6}\right) \\
	&= \frac{3456 Q^2}{L^4 r_h^4} + \frac{41472}{L^6}.
\end{align}

Therefore, the shear viscosity to entropy density ratio to first order in $q_2$ is:
\begin{equation}
	\frac{\eta}{s} = \frac{1}{4\pi} \left[ 1 + q_2 \left( \frac{3456 Q^2}{L^4 r_h^4} + \frac{41472}{L^6} \right) + \mathcal{O}(q_2^2) \right].
\end{equation}

Therefore, for $q_2 > 0$, the KSS bound $\eta/s \geq 1/(4\pi)$ is satisfied, while for $q_2 < 0$ it is violated. As with the conductivity, this suggests that the sign of $q_2$ might be constrained by the requirement of preserving the bound, but again a full analysis of causality and stability is needed to draw definitive conclusions.

In the limit $q_2 \to 0$, we recover the universal value for Einstein gravity:
\begin{equation}
	\frac{\eta}{s} = \frac{1}{4\pi}.
\end{equation}

\subsection*{4.1 Physical Interpretation: Field Theory Coupling}

The ratio of shear viscosity to entropy density has a profound interpretation in the dual field theory. In holography, $\eta/s$ is inversely proportional to the square of the coupling constant $\lambda$ of the boundary theory \cite{Son2007}:

\begin{equation}\label{eta-s-proportionality}
	\frac{\eta}{s} \sim \frac{1}{\lambda^2} \qquad \text{(in appropriate units)}
\end{equation}

where the proportionality constant depends on the specific features of the field theory, such as the number of degrees of freedom. This relation allows us to interpret our results in terms of the strength of interactions in the dual quantum field theory.

For the previously studied non-minimal couplings, we found:

\begin{itemize}
	\item \textbf{$RF^2$ coupling} \cite{Sadeghi:2023hxd}: $\displaystyle \frac{\eta}{s} = \frac{1}{4\pi}$ to first order in $q_2$, implying through Eq.~(\ref{eta-s-proportionality}) that $\lambda$ remains unchanged from the Einstein-Yang-Mills case.
	
	\item \textbf{$R_{\mu\nu\rho\sigma}F^{\mu\nu}F^{\rho\sigma}$ coupling} \cite{Sadeghi:2022bsh}: $\displaystyle \frac{\eta}{s} = \frac{1}{4\pi}$ to first order in $q_2$, again implying no modification to the field theory coupling.
\end{itemize}

In contrast, for the $R^3F^2$ coupling considered in this work, we obtain a non-trivial correction:

\begin{equation}
	\frac{\eta}{s} = \frac{1}{4\pi}\left[1 + q_2\left(\frac{3456Q^2}{L^4r_h^4} + \frac{41472}{L^6}\right) + \mathcal{O}(q_2^2)\right]
\end{equation}

Using the relation (\ref{eta-s-proportionality}), this correction translates directly into a modification of the dual field theory coupling:

\begin{equation}
	\lambda(q_2) = \lambda_0\left[1 - \frac{q_2}{2}\left(\frac{3456Q^2}{L^4r_h^4} + \frac{41472}{L^6}\right) + \mathcal{O}(q_2^2)\right]
\end{equation}

where $\lambda_0$ is the coupling in the Einstein-Yang-Mills case ($q_2 = 0$). Therefore:

\begin{itemize}
	\item For $q_2 > 0$: $\eta/s > 1/4\pi$ $\rightarrow$ the field theory coupling $\lambda$ becomes \textbf{weaker} than in the Einstein-Yang-Mills case.
	
	\item For $q_2 < 0$: $\eta/s < 1/4\pi$ $\rightarrow$ the field theory coupling $\lambda$ becomes \textbf{stronger} than in the Einstein-Yang-Mills case, potentially entering a more strongly coupled regime.
\end{itemize}

This qualitative difference is significant: while the $RF^2$ and $R_{\mu\nu\rho\sigma}F^2$ couplings preserve the KSS bound and leave the field theory coupling unchanged to first order, the $R^3F^2$ coupling provides a new mechanism to \textbf{tune the strength of interactions} in the dual field theory through the gravitational coupling $q_2$. The description of the dual field theory is therefore fundamentally different in our model, as the interaction strength can be continuously varied by the non-minimal coupling parameter.

 \section{Conclusion}

\noindent  In this paper, we have constructed a non-minimal \(R^3 F_{\mu \alpha}^{(a)} F^{(a)\mu \alpha}\) black brane solution in AdS spacetime in four dimensions, treating the \(R^3 F^2\) term as an effective field theory deformation. Due to the complexity of the field equations, we employed a perturbative approach valid to first order in the coupling parameter \(q_2\), under the assumption \(|q_2| / L^6 \ll 1\). Using gauge/gravity duality, we investigated two important holographic transport coefficients: the color non-abelian DC conductivity and the ratio of shear viscosity to entropy density.

Our calculations demonstrate that the non-minimal \(R^3 F^2\) coupling significantly affects these transport coefficients. Specifically, we found that the conductivity bound \(\sigma \geq 1\) is violated for positive values of \(q_2\), while the Kovtun-Son-Starinets bound \(\eta / s \geq 1 / (4\pi)\) is violated for negative \(q_2\). These violations highlight how higher-order non-minimal couplings between curvature and matter fields can modify the transport properties of the dual field theory. From an EFT perspective, such violations may be used to constrain the allowed sign and magnitude of the EFT coupling \(q_2\). However, definitive constraints require a comprehensive analysis of unitarity and causality. Higher-derivative corrections are often associated with ghost-like instabilities that violate unitarity ~\cite{Adams2006}, or with superluminal propagation that violates causality \cite{Camanho2014}. In the holographic context, constraints from CFT unitarity and positivity of energy have been shown to be equivalent to causality in the bulk, and have been successfully applied to constrain higher-derivative couplings in analogous theories ~\cite{Cano2022, Hofman2008, Hofman2009}. Applying such an analysis to the \(R^3 F^2\) coupling would be an important direction for future research, as it would determine whether positive, negative, or both signs of \(q_2\) are physically admissible within a consistent UV completion, and whether the violations of universal bounds we have observed are correlated with fundamental consistency conditions.

In the limit \(q_2 \to 0\), our results smoothly reduce to those of the standard Einstein-Yang-Mills theory, with the conductivity bound saturated and the KSS bound maintaining its universal value of \(1 / (4\pi)\). This consistency check validates our perturbative approach and demonstrates the robustness of our methodology.

The violations of universal bounds observed in our model suggest that non-minimal \(R^3 F^2\) couplings provide a rich framework for exploring beyond-standard-model physics within the holographic context. Future work could extend this analysis to higher-order perturbations, investigate other holographic observables such as Hall conductivity or thermal diffusivity, or study the effects of such couplings on entanglement entropy and other information-theoretic quantities. Additionally, a more systematic exploration of the allowed parameter space from causality and unitarity would further strengthen the EFT interpretation.\\

\vspace{1cm}
\noindent {\large {\bf Acknowledgments} }\\

The author is grateful to the anonymous referee for the thorough review 
of this manuscript and for providing insightful comments and suggestions.\\

\noindent {\large {\bf Data availability statement} }\\

All data that support the findings of this study are included within the article.

\newpage
\appendix
\renewcommand\theequation{\thesection-\arabic{equation}} 
\setcounter{equation}{0}

\appendix
\section{The relation of $\mathcal{L}(t)$}\label{weylsquared}

\begin{equation}
	\mathcal{L}(t) = q_1 \mathcal{A}(t) + q_2 \mathcal{B}(t),
\end{equation}
where,
\begin{align}
	\mathcal{A}(t) &= -t^4 f''(t) - 2t^6 f(t)\left[1 + t H'(t)\right], \\
	\mathcal{B}(t) &= \mathcal{B}_1(t) + \mathcal{B}_2(t) + \mathcal{B}_3(t) + \mathcal{B}_4(t) + \mathcal{B}_5(t),
\end{align}
with,
\begin{align*}
	\mathcal{B}_1(t) &= -4t^4 f''(t)^4 - 4f'(t)^4 (4 - 3t H'(t))^2 \Theta_1(t), \\
	\mathcal{B}_2(t) &= 4t f'(t)^3 (-4 + 3t H'(t)) \left[f''(t)\Theta_2(t) + 3t(4 - 3t H'(t)) f'''(t)\right], \\
	\mathcal{B}_3(t) &= 12t^2 f'(t)^2 f''(t) \left[f''(t)\Theta_3(t) + 2t(-4 + 3t H'(t)) f'''(t)\right],\\
	\mathcal{B}_4(t) &= t^3 f'(t) \left[2q_1 + t H'(t)(q_1 + 76q_2 f''(t)^3) - 4q_2 f''(t)^2(22f''(t) + 3t f'''(t))\right], \\
	\mathcal{B}_5(t) &= -64q_2 f(t)^4 \Gamma(t)^2 \Theta_4(t) \\
	&\quad + 32q_2 f(t)^3 t \Gamma(t) \left\{-t\left[3t\Gamma(t) f'''(t) + f''(t)\Theta_5(t)\right] + f'(t)\Theta_6(t)\right\} \\
	&\quad - 2 f(t) t^3 \left[q_1 t^3 (1 + t H'(t)) - 4q_2\left(f'(t)(-4 + 3t H'(t)) - t f''(t)\right)\right] \times \\
	&\quad \left[t^2 f''(t)\left(f''(t)\Theta_7(t) - 3t f'''(t)\right) + f'(t)^2 \Theta_8(t) - t f'(t)\Theta_9(t)\right] \\
	&\quad - 48q_2 f(t)^2 t^2 \left[t^2 f''(t)\Theta_{10}(t) + f'(t)^2 \Theta_{11}(t) + t f'(t)\Theta_{12}(t)\right],
\end{align*}
and
\begin{align*}
	\Gamma(t) &= (1 - t H'(t))^2 - t^2 H''(t), \\
	\Theta_1(t) &= -14 + t\left[H'(t)(-10 + 9t H'(t)) - 15t H''(t)\right], \\
	\Theta_2(t) &= 28 + 57t H'(t)(-2 + t H'(t)) - 30t^2 H''(t), \\
	\Theta_3(t) &= -38 + t\left[74 H'(t) - 32t H'(t)^2 + 5t H''(t)\right], \\
	\Theta_4(t) &= 7 + t\left[-t H'(t)^2 + t^2 H'(t)^3 - 7 H'(t)(1 + t^2 H''(t)) + t(5 H''(t) + 3t H'''(t))\right], \\
	\Theta_5(t) &= 4 + t\left[8t^2 H'(t)^3 + t^3 H'(t)^4 + H'(t)(14 - 20t^2 H''(t)) \right. \\
	&\quad \left. - t H'(t)^2(27 + 2t^2 H''(t)) + t\left(H''(t)(19 + t^2 H''(t)) + 6t H'''(t)\right)\right], \\
	\Theta_6(t) &= -58 + t\left[13t^3 H'(t)^4 + t^4 H'(t)^5 - t^2 H'(t)^3(53 + 8t^2 H''(t)) \right. \\
	&\quad + t H'(t)^2(17 - 48t^2 H''(t) + 3t^3 H'''(t)) \\
	&\quad + t\left(H''(t)(-49 + 11t^2 H''(t)) - 3t(7 + t^2 H''(t)) H'''(t)\right) \\
	&\quad \left. + H'(t)\left(80 + t^2\left(H''(t)(123 + 7t^2 H''(t)) + 12t H'''(t)\right)\right)\right], \\
	\Theta_7(t) &= -8 + t\left[H'(t)(4 + 3t H'(t)) - 3t H''(t)\right], \\
	\Theta_8(t) &= -20 + t\left[-33t^2 H'(t)^3 + 21t^3 H'(t)^4 - 3t H'(t)^2(24 + 23t^2 H''(t)) \right. \\
	&\quad + 6t\left(H''(t)(7 + 5t^2 H''(t)) - 2t H'''(t)\right) \\
	&\quad \left. + H'(t)\left(118 + 75t^2 H''(t) + 9t^3 H'''(t)\right)\right], \\
	\Theta_9(t) &= 10 + t\left[-51t H'(t)^2 + 30t^2 H'(t)^3 - 2 H'(t)(1 + 18t^2 H''(t)) \right. \\
	&\quad \left. + 3t\left(7 H''(t) + t H'''(t)\right)\right], \\
	\Theta_{10}(t) &= -4 + t\left[3t^2 H'(t)^3 + t^3 H'(t)^4 - 5 H'(t)(-3 + t^2 H''(t)) \right. \\
	&\quad \left. - t H'(t)^2(15 + 2t^2 H''(t)) + t\left(H''(t)(7 + t^2 H''(t)) + t H'''(t)\right)\right], \\
	\Theta_{11}(t) &= 38 + t\left[-t^4 H'(t)^5 + 5t^5 H'(t)^6 - 27t^3 H'(t)^4(3 + t^2 H''(t)) \right. \\
	&\quad + t^2 H'(t)^3(185 + 27t^2 H''(t) + 6t^3 H'''(t)) \\
	&\quad + t\left(H''(t)(55 - t^2 H''(t)(24 + 5t^2 H''(t))) + 8t(1 + t^2 H''(t)) H'''(t)\right) \\
	&\quad + t H'(t)^2\left(-112 + t^2\left(H''(t)(121 + 27t^2 H''(t)) - 11t H'''(t)\right)\right) \\
	&\quad \left. - 2 H'(t)\left(17 + t^2\left(H''(t)(93 + 10t^2 H''(t)) + t(1 + 3t^2 H''(t)) H'''(t)\right)\right)\right], \\
	\Theta_{12}(t) &= -2 + t\left[6t^3 H'(t)^4 - 7t^4 H'(t)^5 + 2t^2 H'(t)^3(26 + 9t^2 H''(t)) \right. \\
	&\quad + 2t\left(17 H''(t) + t(3 + t^2 H''(t)) H'''(t)\right) \\
	&\quad - 2t H'(t)^2\left(50 + t^2\left(H''(t) + t H'''(t)\right)\right) \\
	&\quad \left. - H'(t)\left(-51 + t^2\left(H''(t)(60 + 11t^2 H''(t)) + 2t H'''(t)\right)\right)\right].
\end{align*}


\end{document}